# A violation of the spatial quantum inequality.

Dan Solomon
Rauland-Borg Corporation
Mount Prospect, IL
Email: dan.solomon@rauland.com
Dec. 3, 2010

**Abstract.**

In classical physics the energy density of a field, such as the electromagnetic field, is always positive. However, in quantum field theory it has been shown that the energy density can be negative. There are restrictions, called the quantum inequalities, on the amount of negative energy that can exist in some region of space and time. In this paper we will focus on the spatial quantum inequality as it applies to a massless scalar field in 1-1 dimensional space-time. The spatial quantum inequality is a restriction on the amount of negative energy that can exist in a region of space at a given time. It will be shown that we can specify a quantum state which violates the spatial quantum inequality.

**1. Introduction.**

In classical physics the energy density of a field, such as the electromagnetic field, is always positive. However this is not the case for quantum physics. In quantum field theory the energy density can be negative over some region of space [1]. There are a number of papers that claim to show that there are limits on this effect [2-4]. These limits are called the quantum inequalities. The quantum inequalities provide a lower bound on the weighted average of the energy density over some region of space and time. They apply to free field systems, that is, systems where all external potentials are zero. These have been examined by a number of researchers (see [5] and references, therein). It has speculated that the absence of such limits would result in a violation of the second law of thermodynamics [6] and could give rise to "exotic" phenomenon such as traversable wormholes [7].

Recently a number of papers have been written by this author that claim to demonstrate counterexamples to the quantum inequalities (D. Solomon [8,9,10]). In these papers quantum systems were described which violated the quantum inequalities.

In this paper we will extend the results of Ref. [8] and specify a quantum state which violates the spatial quantum inequality using a different approach from that of Ref. [8].

The spatial quantum inequality applies to a zero mass scalar field in 1-1 dimensional space-time. According to E. E. Flanagen [3] the spatial quantum inequality is expressed by the following equation,

$$\int_{-\infty}^{+\infty} T_{00}(x,t)\rho(x)dx \geq \xi_{S,\min}[\rho] \tag{1.1}$$

where $T_{00}(x,t)$ is the energy density and $\rho(x)$ is a positive weighting function which satisfies,

$$\int_{-\infty}^{+\infty} \rho(x)dx = 1 \tag{1.2}$$

and,

$$\xi_{S,\min}[\rho] = -\frac{1}{24\pi}\int_{-\infty}^{+\infty} dx \frac{\rho'(x)^2}{\rho(x)} \tag{1.3}$$

If we use a Lorentzian sampling function defined by,

$$\rho_L(x) = \frac{\tau}{\pi(x^2 + \tau^2)} \tag{1.4}$$

Then,

$$\xi_{S,\min}[\rho_L] = -\frac{1}{24\pi\tau^2} \tag{1.5}$$

In [8] a system was described that violated this quantum inequality. The analysis in [8] was based on a result by S.G. Mamaev and N.N. Trunov [11] (See also Section 1.6 of [12]). They determined the kinetic energy density of a scalar field with zero mass in 1-1 dimension space-time in the presence of a scalar potential given by,

$$V_\lambda(x) = \lambda\left[\delta(x - a/2) + \delta(x + a/2)\right] \tag{1.6}$$

where $\lambda$ is a non-negative constant. The kinetic energy density is that part of the energy density that is not explicitly dependent on the scalar potential. When the scalar potential is zero the kinetic energy density and energy density are equivalent. Mamaev and Trunov [11] show that for this system the kinetic energy density is given by,

$$T_{00,\lambda}(x) = \begin{cases} -\eta & \text{if } |x| < a/2 \\ 0 & \text{if } |x| > a/2 \end{cases} \tag{1.7}$$

where $\eta$ is a positive constant. Therefore $T_{00,\lambda}(x)$ is negative in the region between $-a/2$ and $+a/2$ and zero elsewhere.

At this point we have a system where the kinetic energy density is completely determined. The quantum inequalities are not applicable for this system because the scalar potential is not zero. This can be easily remedied by instantaneously setting the scalar potential equal to zero. Let us suppose that at $t = 0$ the potential is removed. It was argued in [8] that when this is done the kinetic energy density is continuous with respect to this instantaneous change in the scalar potential. Therefore if $\varepsilon$ is a positive number and $\varepsilon \to 0$ then $T_{00}(x,\varepsilon) \cong T_{00}(x,-\varepsilon) = T_{00,\lambda}(x)$. The result is that we now have a free field system in which the kinetic energy density is given by (1.7) at time $t = \varepsilon \to 0$. Also, at this time, the kinetic energy density is now equivalent to the energy density since the electric potential is now equal to zero. It was shown in [8] that this energy density violates the spatial quantum inequality. This can be demonstrated as follows. Let the weighting function be the Lorentzian as defined in Eq. (1.4) then,

$$\int_{-\infty}^{+\infty} T_{00,\lambda}(x)\rho_L(x)dx = -\eta \int_{-a/2}^{+a/2} \rho_L(x)dx = -\frac{2\eta}{\pi}\arctan\left(\frac{a}{2\tau}\right) \tag{1.8}$$

If the spatial quantum inequality is valid then we can use this result along with (1.5) in (1.1) to obtain,

$$-\frac{2\eta}{\pi}\arctan\left(\frac{a}{2\tau}\right) \geq -\frac{1}{24\pi\tau^2} \tag{1.9}$$

In the limit that $\tau \gg a$, so that $a/2\tau$ is small, use $\arctan(a/2\tau) \cong a/2\tau$ in the above to obtain,

$$-\frac{2\eta}{\pi}\frac{a}{2\tau} \underset{\tau \gg a}{\geq} -\frac{1}{24\pi\tau^2} \tag{1.10}$$

From this we obtain $-\eta a \underset{\tau \gg a}{\geq} -1/24\tau$ which yields $1/24\tau \underset{\tau \gg a}{\geq} \eta a$. This last equality is not true for sufficiently large $\tau$. As $\tau \to \infty$, Eq. (1.10) becomes $0 \underset{\tau \to \infty}{\geq} \eta a$ which is obviously false because the right hand side is positive. Therefore the spatial quantum inequality is violated in this situation.

The above result depends on the argument that the change in the kinetic energy density is continuous when there is an abrupt change in the scalar potential. In [8] this was shown to be the case based on an analysis of this system using the Heisenberg picture formulation of quantum field theory. In the next section we will show that this argument is also consistent with an analysis based on the Schrödinger picture. Following this it will be shown that the "artifact" of instantaneously removing the scalar potential is unnecessary and we will directly specify a quantum state in a free field which violates the spatial quantum inequality.

**2. Analysis in the Schrödinger picture.**

Consider a zero mass scalar field in 1-1 dimensional space-time in the Schrödinger picture. In the Schrödinger picture the time dependence of the system is associated with the state vector $|\Omega(t)\rangle$ and the field operators $\hat{\varphi}(x)$ and $\hat{\pi}(x)$ are time independent. The state vector evolves in time according to the equation,

$$i\frac{\partial |\Omega(t)\rangle}{\partial t} = \hat{H}|\Omega(t)\rangle \tag{2.1}$$

where the Hamiltonian operator $\hat{H}$ is given by,

$$\hat{H} = \hat{H}_0 + \frac{1}{2}\int V(x,t)\hat{\varphi}(x)\cdot\hat{\varphi}(x)dx \tag{2.2}$$

where $\hat{H}_0$ is the free field Hamiltonian operator and is given by,

$$\hat{H}_0 = \int \hat{T}_{00}(x)dx \tag{2.3}$$

and where $\hat{T}_{00}(x)$ is the kinetic energy density operator which is given by,

$$\hat{T}_{00}(x) = \frac{1}{2}\left(\hat{\pi}(x)\cdot\hat{\pi}(x) + \frac{d\hat{\varphi}(x)}{dx}\cdot\frac{d\hat{\varphi}(x)}{dx}\right) \tag{2.4}$$

The *total* energy density operator is,

$$\hat{\xi}_{00}(x) = \hat{T}_{00}(x) + \frac{1}{2}V(x,t)\hat{\varphi}(x)\cdot\hat{\varphi}(x) \tag{2.5}$$

When the scalar potential is zero the kinetic energy density and energy density operators are equivalent.

The field operators $\hat{\varphi}(x)$ and $\hat{\pi}(x)$ obey the commutation relationship,

$$[\hat{\varphi}(g_1), \hat{\pi}(g_2)] = i\int g_1 g_2 \, dx \tag{2.6}$$

with all other commutations being zero. In the above expression,

$$\hat{\varphi}(g_1) = \int \hat{\varphi}(x) g_1(x) \, dx \text{ and } \hat{\pi}(g_2) = \int \hat{\pi}(x) g_2(x) \, dx \tag{2.7}$$

where $g_1(x)$ and $g_2(x)$ are arbitrary functions.

If $|\Omega(t)\rangle$ is a normalized state vector then the kinetic energy density expectation value is given by,

$$T_{00}(x,t;|\Omega(t)\rangle) = \langle \Omega(t) | \hat{T}_{00}(x) | \Omega(t) \rangle \tag{2.8}$$

There are some problems associated with evaluating this expression. These are due to the fact that the quantities $\hat{\pi}(x) \cdot \hat{\pi}(x)$ and $\frac{\partial \hat{\varphi}(x)}{\partial x} \cdot \frac{\partial \hat{\varphi}(x)}{\partial x}$ are ill-defined and highly divergent. As a result of this $\langle \Omega(t) | \hat{T}_{00}(x) | \Omega(t) \rangle$ will be infinite. A possible way to resolve this problem is to subtract off a quantity that corresponds to the kinetic energy density of the "unperturbed" vacuum state. Therefore the "renormalized" kinetic energy density is given by,

$$T_{00,R}(x;|\Omega(t)\rangle) = \langle \Omega(t) | \hat{T}_{00}(x) | \Omega(t) \rangle - T_{00,vac} \tag{2.9}$$

where $T_{00,vac}$ is a renormalization constant. Note that this is a formal expression only, due to the fact that both quantities on the right of the equation are infinite.

Let the scalar potential satisfy,

$$V(x,t) = \theta(-t) V(x) \tag{2.10}$$

Therefore Eq. (2.1) can be written as,

$$i\frac{\partial |\Omega^{(-)}(t)\rangle}{\partial t} = \left( \hat{H}_0 + \frac{1}{2} \int V(x) \hat{\varphi}(x) \cdot \hat{\varphi}(x) \right) |\Omega^{(-)}(t)\rangle = \text{ for } t < 0 \tag{2.11}$$

and,

$$i\frac{\partial |\Omega^{(+)}(t)\rangle}{\partial t} = \hat{H}_0 |\Omega^{(+)}(t)\rangle \text{ for } t \geq 0 \tag{2.12}$$

where the state vector $|\Omega(t)\rangle$ is represented by $|\Omega^{(-)}(t)\rangle$ for $t<0$ and $|\Omega^{(+)}(t)\rangle$ for $t\geq 0$. Since this is first order differential equation the boundary condition, at $t=0$, is,

$$|\Omega^{(+)}(t)\rangle = |\Omega^{(-)}(t)\rangle \tag{2.13}$$

Use this in (2.9) to obtain $T_{00,R}\left(x;|\Omega^{(+)}(t)\rangle\right) = T_{00,R}\left(x;|\Omega^{(-)}(t)\rangle\right)$. Therefore the kinetic energy density is continuous at $t=0$. This confirms the result of Ref [8] which obtained the same result working in the Heisenberg picture.

In the rest of the paper we will show that it is possible to demonstrate a violation of the spatial quantum inequality without requiring the "artifact" of an instantaneous removal of the scalar potential. The basic outline of the argument is as follows. Mamaev and Trunov [11] examined a system in which the kinetic energy density violates the spatial quantum inequality. This system is represented by a particular normalized state vector which we define by $|0_\lambda\rangle$. Referring to (2.9) a formal expression for the kinetic energy density for this state vector is,

$$T_{00,R}\left(x;|0_\lambda\rangle\right) = \langle 0_\lambda|\hat{T}_{00}(x)|0_\lambda\rangle - T_{00,vac} \tag{2.14}$$

Mamaev and Trunov [11] formulated their problem in the presence of the non-zero scalar potential given by (1.6). However the state vector $|0_\lambda\rangle$ is a mathematical object that is independent of the scalar potential. Therefore if the quantum inequalities hold for all state vectors they should hold for the state vector $|0_\lambda\rangle$ also. However according to the calculation carried out in Section 1 this is not the case. This will be explored further in the rest of the paper.

**3. Mode expansion of the field operators.**

The field operators must satisfy the commutation relationships in Eq. (2.6). Let $\hat{\varphi}(x)$ and $\hat{\pi}(x)$ be given by,

$$\hat{\varphi}(x) = \sum_n \left(\hat{a}_n u_n(x) + \hat{a}_n^* u_n^*(x)\right) \tag{3.1}$$

and,

$$\hat{\pi}(x) = \sum_n \left(\hat{a}_n v_n(x) + \hat{a}_n^* v_n^*(x)\right) \tag{3.2}$$

Note that we are working in the Schrödinger picture so that field operators are time independent. We will assume that the system is in a box of length $L$ where $L \to \infty$. In this case the boundary conditions on $u_n$ and $v_n$ are,

$$u_n(\pm L/2) = 0 \text{ and } v_n(\pm L/2) = 0 \tag{3.3}$$

In the above expressions the quantities $u_n(x)$ and $v_n(x)$ may be thought of as expansion functions and the $\hat{a}_n$ and $\hat{a}_n^*$ may be thought of as expansion coefficients. These quantities must be defined in such a way that the commutations (2.6) are obeyed. This will be the case if the expansion functions $u_n(x)$ and $v_n(x)$ satisfy,

$$(u_n, v_m) = \delta_{mn} \text{ and } (u_n^*, v_m) = 0 \tag{3.4}$$

where the scalar product is defined by,

$$(u_n, v_m) \equiv -i \int_{-L/2}^{+L/2} (u_n v_m^* - v_n u_m^*) dx; \quad (u_n^*, v_m) \equiv -i \int_{-L/2}^{+L/2} (u_n^* v_m^* - v_n^* u_m^*) dx \tag{3.5}$$

(Note – one must be careful in using these expressions. For example, if $u_n$ is real the relationship $(u_n, v_m) = (u_n^*, v_m)$ will, in general, not be true because $v_n$ may not be real.) The expansion coefficients, $\hat{a}_n$, are given by,

$$\hat{a}_n = -i \int_{-L/2}^{+L/2} (\hat{\varphi} v_n^* - \hat{\pi} u_n^*) dx \tag{3.6}$$

It can be shown that the $\hat{a}_n$ and $\hat{a}_n^*$ satisfy the usual commutation relationship,

$$[\hat{a}_m, \hat{a}_n^*] = \delta_{nm} \tag{3.7}$$

with all other commutations being zero.

Eq. (3.4) is satisfied if the $u_n$ are mode solutions of the eigenvalue equation,

$$-\frac{\partial^2 u_n(x)}{\partial x^2} + U(x) u_n(x) = \omega_n^2 u_n(x) \tag{3.8}$$

and the $v_n$ are defined by the relationship,

$$v_n = -i\omega_n u_n \tag{3.9}$$

where $U(x)$ is an arbitrary function and the $\omega_n$ are the eigenvalues. Using (3.4), (3.5), and (3.8) we obtain,

$$(u_n, v_n) = 2\omega_n \int_{-L/2}^{+L/2} u_n u_n^* dx = 1 \tag{3.10}$$

which is the normalization condition on the mode solutions $u_n$.

In formulating the field operators we note that the mode solutions $u_n(x)$ are not unique but depend on the selection of $U(x)$. So how do we determine the proper $U(x)$? That will be considered later but for the moment let us set $U(x) = 0$ in (3.8). For this case the mode solutions will be designated by $u_{0,jn}$ and $v_{0,jn}$ where $j = 1$ or $2$ and $n = 1, 2, 3, \ldots$. The $u_{0,jn}$ are given by,

$$u_{0,1n}(x) = \frac{\sin(\omega_{0,1n} x)}{\sqrt{\omega_{0,1n} L}} \text{ and } u_{0,2n}(x) = \frac{\cos(\omega_{0,2n} x)}{\sqrt{\omega_{0,2n} L}} \tag{3.11}$$

In order to satisfy the boundary conditions (3.3) the eigenvalues $\omega_{0,jn}$ are given by,

$$\omega_{0,1n} = \frac{2\pi n}{L} \text{ and } \omega_{0,2n} = \frac{2\pi}{L}\left(n - \frac{1}{2}\right) \tag{3.12}$$

From (3.9) the $v_{0,jn}$ are given by $v_{0,jn} = -i\omega_{0,jn} u_{0,jn}$.

For this case the expression for the field operators is given by,

$$\hat{\varphi}(x) = \sum_{jn}\left(\hat{a}_{0,jn} u_{0,jn} + \hat{a}_{0,jn}^* u_{0,jn}^*\right) \tag{3.13}$$

and,

$$\hat{\pi}(x) = -i\sum_{jn} \omega_{0,jn}\left(\hat{a}_{0,jn} u_{0,jn} - \hat{a}_{0,jn}^* u_{0,jn}^*\right) \tag{3.14}$$

We will define the state vector for the vacuum state $|0\rangle$ as the normalized state vector which is destroyed by all annihilation operators $\hat{a}_{0,jn}$, that is,

$$\hat{a}_{0,jn}|0\rangle = 0 \tag{3.15}$$

We will define renormalization constant $T_{00,vac}$ by the expression,

$$T_{00,vac} = \langle 0|\hat{T}_{00}(x)|0\rangle \tag{3.16}$$

Note that this is a formal expression, at this point, because it is infinite. However one direct consequence of this definition is that $T_{00,R}(x;|0\rangle)$, the renormalized kinetic energy density of the vacuum state, is zero.

Next consider the case where $U(x) = V_\lambda(x)$. Recall that $V_\lambda(x)$ is defined by (1.7). For this case the expansion functions will be designated by $u_{\lambda,jn}$ and $v_{\lambda,jn}$. Using this set of expansion functions, along with $v_{\lambda,jn} = -i\omega_{\lambda,jn} u_{\lambda,jn}$, the field operators are expressed as,

$$\hat{\varphi}(x) = \sum_{jn} \left( \hat{a}_{\lambda,jn} u_{\lambda,jn} + \hat{a}^*_{\lambda,jn} u^*_{\lambda,jn} \right) \tag{3.17}$$

and,

$$\hat{\pi}(x) = -i \sum_{jn} \omega_{\lambda,jn} \left( \hat{a}_{\lambda,jn} u_{\lambda,jn} - \hat{a}^*_{\lambda,jn} u^*_{\lambda,jn} \right) \tag{3.18}$$

where the $\hat{a}_{\lambda,jn}$ are obtained by using the mode solutions $u_{\lambda,jn}$ and $v_{\lambda,jn}$ in (3.6).

The normalized state vector $|0_\lambda\rangle$ is defined by,

$$\hat{a}_{\lambda,jn} |0_\lambda\rangle = 0 \text{ for all } \hat{a}_{\lambda,jn} \tag{3.19}$$

Note that in defining the state vectors $|0_\lambda\rangle$ and $|0\rangle$ we have not made any statements about the scalar potential for the systems in which these state vectors exist. For example $|0_\lambda\rangle$ could be the state vector for a system in which the scalar potential is zero or $V_\lambda(x)$ or any other function. The same is true for $|0\rangle$. However if the scalar potential of the system is zero then the state vector $|0\rangle$ with be the lowest energy state for that system. Similarly if the scalar potential is $V_\lambda(x)$ then $|0_\lambda\rangle$ will be lowest energy state for that system. Essentially $|0\rangle$ and $|0_\lambda\rangle$ are the *vacuum states* for a system where the scalar potential is zero and $V_\lambda(x)$, respectively. This is why Mamaev and Trunov [11] use the state $|0_\lambda\rangle$ for their analyses of a system where the scalar potential is given by $V_\lambda(x)$. They want to use the lowest energy state. However there is no reason why the state vector $|0_\lambda\rangle$ could not exist in a system where the scalar potential is zero or any other value.

## 4. The kinetic energy density.

In this section we will find an expression for the kinetic energy density of the state $|0_\lambda\rangle$. From the above discussion the formal expression for the kinetic energy density is,

$$T_{00,R}(x;|0_\lambda\rangle) = \langle 0_\lambda|\hat{T}_{00}(x)|0_\lambda\rangle - \langle 0|\hat{T}_{00}(x)|0\rangle \qquad (4.1)$$

To evaluate this we will start by writing an expression for the quantity $\langle 0_\lambda|\hat{T}_{00}(x)|0_\lambda\rangle$. Using (2.4) we can write this as,

$$\langle 0_\lambda|\hat{T}_{00}(x)|0_\lambda\rangle = \frac{1}{2}\langle 0_\lambda|\left(\hat{\pi}(x)\cdot\hat{\pi}(x) + \frac{\partial\hat{\varphi}(x)}{\partial x}\cdot\frac{\partial\hat{\varphi}(x)}{\partial x}\right)|0_\lambda\rangle \qquad (4.2)$$

The first problem that we must address is related to the fact that we have specified two different equations for the field operators. In Eqs. (3.13) and (3.14) the field operators are expanded in terms of the mode solutions $u_{0,jn}$ and in Eqs. (3.17) and (3.18) the field operators are expanded in terms of the mode solutions $u_{\lambda,jn}$. Which mode solutions should we use in the expansion of the field operator? It makes sense to use the mode solutions that simplify the problem. We already obtained a simple relation between $\hat{a}_{\lambda,jn}$ and $|0_\lambda\rangle$. On the other hand the relation between $\hat{a}_{0,jn}$ and $|0_\lambda\rangle$ will be quite complicated. Therefore we will evaluate $\langle 0_\lambda|\hat{T}_{00}(x)|0_\lambda\rangle$ using the form of the field operators given by Eqs. (3.17) and (3.18). Doing this we obtain,

$$\langle 0_\lambda|\hat{T}_{00}(x)|0_\lambda\rangle = \sum_{jn} T_{\lambda,jn}(x) \qquad (4.3)$$

where $T_{\lambda,jn}(x)$ is the kinetic energy density associated with the mode $jn$ and is given by,

$$T_{\lambda,jn}(x) = \frac{1}{2}\left(|\omega_{\lambda,jn} u_{\lambda,jn}|^2 + \left|\frac{\partial u_{\lambda,jn}}{\partial x}\right|^2\right) \qquad (4.4)$$

Next we determine the kinetic energy density for the vacuum state $|0\rangle$. In this case we use the form of the field operator given by (3.13) and (3.14) to obtain,

$$T_{00,vac}(x) = \langle 0|\hat{T}_{00}(x)|0\rangle = \sum_{jn} T_{0,jn}(x) \qquad (4.5)$$

where,

$$T_{0,jn}(x) = \frac{1}{2}\left(\left|\omega_{0,jn}u_{0,jn}\right|^2 + \left|\frac{\partial u_{0,jn}}{\partial x}\right|^2\right) \tag{4.6}$$

Using (3.11) we obtain,

$$T_{0,jn}(x) = \frac{\omega_{0,jn}}{2L} \tag{4.7}$$

Note that the $T_{0,jn}(x)$ are independent of $x$. From the above discussion we can write the kinetic energy density as,

$$T_{00,R}(x;|0_\lambda\rangle) = \sum_{jn}T_{\lambda,jn}(x) - \sum_{jn}T_{0,jn} \tag{4.8}$$

This is still a formal expression because each of the summations is infinite therefore we are subtracting one infinity from another. The way this was handled by Mamaev and Trunov [11] was to combine the expressions under the one summation sign to obtain,

$$T_{00,R}(x;|0_\lambda\rangle) = \sum_{jn}\left(T_{\lambda,jn}(x) - T_{0,jn}\right) \tag{4.9}$$

This is called "mode renormalization". This will yield a finite result because the quantity in parenthesis fall off sufficiently fast as the index $n \to \infty$. It will then be shown in Section 6 that mode renormalization yields a physically sensible solution.

**5. Calculation of the kinetic energy density.**

In this section we will determine $T_{00,R}(x;|0_\lambda\rangle)$ as specified in (4.9). In order to evaluate this we must write down the mode solutions $u_{\lambda,jn}$. Define region I as the region where $|x| < (a/2)$ and region II as the region $|x| > (a/2)$. From (3.8) and using $V(x) = V_\lambda(x)$ we obtain,

$$\frac{\partial^2 u_{\lambda,jn}(x;I)}{\partial x^2} + \omega^2_{\lambda,jn}u_{\lambda,jn}(x;I) = 0 \text{ for } |x| < a/2 \tag{5.1}$$

and,

$$\frac{\partial^2 u_{\lambda,jn}(x;II)}{\partial x^2} + \omega^2_{\lambda,jn}u_{\lambda,jn}(x;II) = 0 \text{ for } |x| > a/2 \tag{5.2}$$

The boundary condition at $x = \pm a/2$, due to the delta function potential, are,

$$u_{\lambda,jn}(\pm a/2;I) = u_{\lambda,jn}(\pm a/2;II) \tag{5.3}$$

and,

$$\frac{\partial u_{\lambda,jn}(a/2;II)}{\partial x} - \frac{\partial u_{\lambda,jn}(a/2;I)}{\partial x} = \lambda u_{\lambda,jn}(a/2) \quad (5.4)$$

along with,

$$\frac{\partial u_{\lambda,jn}(-a/2;I)}{\partial x} - \frac{\partial u_{\lambda,jn}(-a/2;II)}{\partial x} = \lambda u_{\lambda,jn}(-a/2) \quad (5.5)$$

We solve these equations for the mode solutions $u_{\lambda,jn}$ by using the results from [11], along with the boundary condition $u_{\lambda,jn}(\pm L/2) = 0$, and the normalization condition (3.10) to obtain,

$$u_{\lambda,1n}(x) = \frac{N_1(\omega_{\lambda,1n})}{\sqrt{\omega_{\lambda,1n}L}} \times \begin{cases} A_1(\omega_{\lambda,1n})\sin(\omega_{\lambda,1n}x), & |x| < a/2 \\ \sin[\omega_{\lambda,1n}x + \tau(x)\delta_1(\omega_{\lambda,1n})], & |x| > a/2 \end{cases} \quad (5.6)$$

$$u_{\lambda,2n}(x) = \frac{N_2(\omega_{\lambda,2n})}{2\sqrt{\omega_{\lambda,2n}L}} \times \begin{cases} A_2(\omega_{\lambda,2n})\cos(\omega_{\lambda,2n}x), & |x| < a/2 \\ \cos(\omega_{\lambda,2n}x + \delta_2(\omega_{\lambda,2n})\tau(x)), & |x| > a/2 \end{cases} \quad (5.7)$$

where,

$$\omega_{\lambda,jn} = \omega_{0,jn} - \frac{2\delta_j(\omega_{\lambda,jn})}{L} \quad (5.8)$$

with,

$$A_1(\omega_{\lambda,1n}) = \left[\sin^2\Omega_{\lambda,1n} + \left((\Lambda/\Omega_{\lambda,1n})\sin\Omega_{\lambda,1n} + \cos\Omega_{\lambda,1n}\right)^2\right]^{-(1/2)} \quad (5.9)$$

and,

$$A_2(\omega_{\lambda,2n}) = \left[\cos^2\Omega_{\lambda,2n} + \left((\Lambda/\Omega_{\lambda,2n})\cos\Omega_{\lambda,2n} - \sin\Omega_{\lambda,2n}\right)^2\right]^{-(1/2)} \quad (5.10)$$

and

$$\delta_1(\omega_{\lambda,1n}) = \tan^{-1}\left\{\frac{-(\Lambda/\Omega_{\lambda,1n})\sin^2\Omega_{\lambda,1n}}{(1+(\Lambda/2\Omega_{\lambda,1n})\sin(2\Omega_{\lambda,1n}))}\right\} \quad (5.11)$$

$$\delta_2(\omega_{\lambda,2n}) = \tan^{-1}\left\{\frac{-(\Lambda/\Omega_{\lambda,2n})\cos^2\Omega_{\lambda,2n}}{(1-(\Lambda/2\Omega_{\lambda,2n})\sin(2\Omega_{\lambda,2n}))}\right\} \quad (5.12)$$

with,

$\Lambda = \lambda a/2$, $\Omega_{\lambda,jn} = \omega_{\lambda,jn}a/2$, and $\tau(x) = +1$ for $x > 0$ and $\tau(x) = -1$ for $x < 0$.

In order to simplify notation we often write $A_{jn}$ instead of $A_j(\omega_{\lambda,jn})$ so that the dependence of $A_j$ on $\omega_{\lambda,jn}$ is indicated by the index $n$. Similarly we may write for $N_{jn}$ for $N_j(\omega_{\lambda,jn})$ and $\delta_{jn}$ for $\delta_j(\omega_{\lambda,jn})$. The same will be true for the function $B_j(\omega_{\lambda,jn})$ introduced below.

The $N_{jn}$ are chosen to satisfy the normalization condition (3.10) which yields,

$$N_{jn} = 1\bigg/\sqrt{1-\frac{B_{jn}}{L}} \tag{5.13}$$

where,

$$B_{1n} = B_1(\omega_{\lambda,1n}) = a(1-A_{1n}^2) + \frac{A_{1n}^2 \sin(\omega_{\lambda,1n}a) - \sin(\omega_{\lambda,1n}a + 2\delta_{1n})}{\omega_{\lambda,1n}} \tag{5.14}$$

and,

$$B_{2n} = B_2(\omega_{\lambda,2n}) = a(1-A_{2n}^2) - \frac{A_{2n}^2 \sin(\omega_{\lambda,2n}a) - \sin(\omega_{\lambda,2n}a + 2\delta_{2n})}{\omega_{\lambda,2n}} \tag{5.15}$$

Use these relationships in (4.9) along with (4.6) and (4.7) to obtain for region I,

$$T_{00,R}(x,I;|0_\lambda\rangle) = \frac{1}{2L}\sum_{jn}\left(N_{jn}^2 A_{jn}^2 \omega_{\lambda,jn} - \omega_{0,jn}\right) \tag{5.16}$$

Similarly, for region II we obtain,

$$T_{00,R}(x,II;|0_\lambda\rangle) = \frac{1}{2L}\sum_{jn}\left(\omega_{\lambda,jn}N_{jn}^2 - \omega_{0,jn}\right) \tag{5.17}$$

Calculate $T_{00,R}(x,I;|0_\lambda\rangle)$ first. In the limit $L\to\infty$ we can write,

$$\sum_{jn} \to \frac{L}{2\pi}\int \sum_{j=1,2} d\omega_{0,j} \tag{5.18}$$

$$T_{00,R}(x,I;|0_\lambda\rangle) = \frac{1}{4\pi}\int_0^\infty \left\{\sum_{j=1,2}\left(\omega_{\lambda,j}N_j^2(\omega_{\lambda,j})A_j^2(\omega_{\lambda,j}) - \omega_{0,j}\right)d\omega_{0,j}\right\} \tag{5.19}$$

where, from (5.8),

$$\omega_{\lambda,j} = \omega_{0,j} - \frac{2\delta_j(\omega_{\lambda,j})}{L} \tag{5.20}$$

(Note we have reverted back to our original notation where the explicit dependence on the $\omega_{\lambda,j}$ is indicated because in the integral the index $n$ is not used)

Since the integrand falls of sufficiently fast for large $\omega_{0,j}$ we can write,

$$T_{00,R}\left(x,I;|0_\lambda\rangle\right) \underset{S\to\infty}{=} \frac{1}{4\pi}\int_0^S \left\{\sum_{j=1,2}\left(\omega_{\lambda,j}N_j^2(\omega_{\lambda,j})A_j^2(\omega_{\lambda,j})-\omega_{0,j}\right)d\omega_{0,j}\right\} \quad (5.21)$$

We can write this as,

$$T_{00,R}\left(x,I;|0_\lambda\rangle\right) \underset{S\to\infty}{=} \delta T_{00,R}\left(x,I;|0_\lambda\rangle\right) - \frac{1}{2\pi}\int_0^S \omega\, d\omega \quad (5.22)$$

where we have used,

$$\int_0^S\left(\sum_{j=1,2}\omega_{0,j}d\omega_{0,j}\right) = 2\int_0^S \omega\, d\omega \quad (5.23)$$

and where,

$$\delta T_{00,R}\left(x,I;|0_\lambda\rangle\right) = \frac{1}{4\pi}\int_0^S\left\{\sum_{j=1,2}N_{\lambda,j}^2(\omega_{\lambda,j})A_j^2(\omega_{\lambda,j})\omega_{\lambda,j}d\omega_{0,j}\right\} \quad (5.24)$$

From (5.20) we obtain,

$$d\omega_{0,j} = d\omega_{\lambda,j}\left(1+\frac{2}{L}\frac{d\delta_j(\omega_{\lambda,j})}{d\omega_{\lambda,j}}\right) \quad (5.25)$$

Use this result in (5.19) to obtain,

$$\delta T_{00}\left(x,I;|0_\lambda\rangle\right) = \frac{1}{4\pi}\int_0^S\sum_{j=1,2}N_j^2(\omega_{\lambda,j})A_j^2(\omega_{\lambda,j})\omega_{\lambda,j}\left(1+\frac{2}{L}\frac{d\delta_j(\omega_{\lambda,j})}{d\omega_{\lambda,j}}\right)d\omega_{\lambda,j} \quad (5.26)$$

Also note that in the limit $L\to\infty$,

$$N_j^2(\omega) = 1+\frac{B_j(\omega)}{L}+O(1/L^2) \quad (5.27)$$

Drop terms that are $O(1/L)$ to obtain,

$$\delta T_{00}\left(x,I;|0_\lambda\rangle\right) = \frac{1}{4\pi}\int_0^S\left(A_1^2(\omega)+A_2^2(\omega)\right)d\omega \quad (5.28)$$

Use this in (5.22) to obtain,

$$T_{00}\left(x,I;|0_\lambda\rangle\right) \underset{S\to\infty}{=} \frac{1}{4\pi}\int_0^S\left(A_{1n}^2(\omega)+A_{2n}^2(\omega)-2\omega\right)d\omega \quad (5.29)$$

Using the results of Ref. [11] this can be rewritten as,

$$T_{00,R}(x,I;|0_\lambda\rangle) = \eta_1 + \eta_2 \tag{5.30}$$

$$\eta_1 = -\frac{\Lambda}{\pi a^2}\int_0^\infty \frac{ye^{-y}dy}{ye^y + \Lambda\sinh y}; \quad \eta_2 = \frac{\Lambda}{\pi a^2}\int_0^\infty \frac{ye^{-y}dy}{ye^y + \Lambda\cosh y} \tag{5.31}$$

with $\Lambda = \lambda a/2$. It can be shown that $\eta_1 + \eta_2 < 0$ based on the fact that $\cosh y \geq \sinh y$. Therefore in the region I the energy density is negative and independent of $x$.

Similarly, for region II we obtain,

$$T_{00,R}(x,II;|0_\lambda\rangle) = \frac{1}{4\pi}\int_0^\infty \sum_{j=1,2}\left\{\omega\left(1 + \frac{B_j(\omega)}{L} + O(1/L^2)\right)\left(1 + \frac{2}{L}\frac{d\delta_{j\omega}}{d\omega}\right) - \omega\right\}d\omega \tag{5.32}$$

This becomes,

$$T_{00,R}(x,II;|0_\lambda\rangle) = \frac{\beta}{L} + O(1/L^2) \tag{5.33}$$

where,

$$\beta = \frac{1}{4\pi}\int_0^\infty \sum_{j=1,2}\left\{\omega_j\left(B_j(\omega_j) + 2\frac{d\delta_j(\omega_j)}{d\omega_j}\right)\right\}d\omega_j \tag{5.34}$$

It can be shown that $\beta$ is a finite number. Since we are operating in the limit $L\to\infty$ we obtain,

$$T_{00,R}(x,II;|0_\lambda\rangle)\underset{L\to\infty}{=} 0 \tag{5.35}$$

The result of this is that the kinetic energy density of the state $|0_\lambda\rangle$ is negative in region I and zero in region II. As is shown in the Introduction this violates the spatial quantum inequality.

One potential concern with the above result that it seems to indicate the total kinetic energy density integrated over all space is negative. This because the contribution to the integral from region I is $-|(\eta_1 + \eta_2)|a$ and, if we refer to Eq. (5.35), the contribution from region II is zero. The result is that the total integrated kinetic energy is negative. This is, of course, not possible because the total kinetic energy must be positive. This can be explained by noting that we obtained (5.35) by taking the limit $L\to\infty$. If we integrate the kinetic energy density over all space the contribution from region II is actually $(\beta/L + O(1/L^2)) \cdot (L-a)$ which equals $\beta$ in the limit that $L \to \infty$. The result is that the total kinetic energy is $\beta - \eta a$ which can be shown to be positive.

## 6. Confirmation of mode renormalization.

The results of the last section were based on the use of mode renormalization. The question that we must address is whether or not mode renormalization yields a physical correct result or whether some other form of renormalization should be used instead. In this section we will compare these results against an alternative calculation which will be shown to be renormalization independent. It will be shown that both methods yield identical results.

Refer back to (2.9) and note that the infinite renormalization constant $T_{00,vac}$ is independent of $x$. If we take the derivative of (2.9) with respect to $x$ we obtain,

$$\frac{d}{dx}T_{00,R}(x;|\Omega\rangle) = \frac{d}{dx}\langle\Omega|\hat{T}_{00}(x)|\Omega\rangle \tag{6.1}$$

Note that the renormalization constant is gone. The result of this is that even though $\langle\Omega|\hat{T}_{00}(x)|\Omega\rangle$ is infinite and undefined it is reasonable to expect that the derivative of $\langle\Omega|\hat{T}_{00}(x)|\Omega\rangle$ is finite and well defined.

Another way to look at this is to consider the kinetic density operator $\hat{T}_{00}(x)$. As stated previously the problem with evaluating this operator is that quantities such as $\langle 0|\hat{\pi}(x) \cdot \hat{\pi}(x)|0\rangle$ are ill-defined and divergent. Let $\hat{o}(x)$ stand for the operator $\hat{\pi}(x)$, $\hat{\varphi}(x)$, or $d\hat{\varphi}(x)/dx$. The quantity $\langle 0|\hat{o}(x) \cdot \hat{o}(x)|0\rangle$ is ill-defined and divergent but the quantity $\langle 0|\hat{o}(x+\varepsilon) \cdot \hat{o}(x)|0\rangle$ for $\varepsilon \neq 0$ will be well-defined and is finite. Next consider the quantity $\langle 0|d\hat{o}(x)/dx \cdot \hat{o}(x)|0\rangle$. By the normal definition of the derivative this can be written as,

$$\langle 0|\frac{d\hat{o}(x)}{dx} \cdot \hat{o}(x)|0\rangle \underset{\varepsilon \to 0}{=} \langle 0|\frac{\hat{o}(x+\varepsilon) - \hat{o}(x-\varepsilon)}{\varepsilon} \cdot \hat{o}(x)|0\rangle \tag{6.2}$$

Rearrange terms to obtain,

$$\langle 0|\frac{d\hat{o}(x)}{dx} \cdot \hat{o}(x)|0\rangle \underset{\varepsilon \to 0}{=} \frac{\langle 0|\hat{o}(x+\varepsilon) \cdot \hat{o}(x)|0\rangle - \langle 0|\hat{o}(x-\varepsilon) \cdot \hat{o}(x)|0\rangle}{\varepsilon} \tag{6.3}$$

Since each term is well defined one should expect that the derivative is well defined.

Based on the above analysis we will assume that the quantity $\langle\Omega|d\hat{T}_{00}(x)/dx|\Omega\rangle$ can be evaluated without the use of renormalization. This means that the change in the kinetic energy density from $x_1$ to $x_2$ can be determined without using renormalization and is given by,

$$\Delta T_{00}\left(x_1 \to x_2; |\Omega\rangle\right) = \int_{x_1}^{x_2} \langle\Omega|\frac{d\hat{T}_{00}(x)}{dx}|\Omega\rangle dx \tag{6.4}$$

where,

$$\frac{d\hat{T}_{00}(x)}{dx} = \frac{1}{2}\left(\begin{array}{c}\frac{d\hat{\pi}(x)}{dx}\cdot\hat{\pi}(x) + \hat{\pi}(x)\cdot\frac{d\hat{\pi}(x)}{dx} \\ + \frac{\partial^2\hat{\varphi}(x)}{\partial x^2}\cdot\frac{\partial\hat{\varphi}(x)}{\partial x} + \frac{\partial\hat{\varphi}(x)}{\partial x}\cdot\frac{\partial^2\hat{\varphi}(x)}{\partial x^2}\end{array}\right) \tag{6.5}$$

Let's evaluate $\Delta T_{00}\left(x_1 \to x_2; |0_\lambda\rangle\right)$. Using (4.5) and (4.6) and noting that for our problem the $u_{\lambda,jn}$ are real we obtain,

$$\langle 0_\lambda|\frac{d\hat{T}_{00}(x)}{dx}|0_\lambda\rangle = \sum_{jn}\left(\omega_{\lambda,jn}^2 u_{\lambda,jn}\cdot\frac{du_{\lambda,jn}}{dx} + \frac{du_{\lambda,jn}}{dx}\cdot\frac{d^2 u_{\lambda,jn}}{dx^2}\right) \tag{6.6}$$

Use (3.8) with $U(x) = V_\lambda(x)$ in the above to obtain,

$$\langle 0_\lambda|\frac{d\hat{T}_{00}(x)}{dx}|0_\lambda\rangle = \sum_{jn}\left(V_\lambda(x)u_{\lambda,jn}(x)\cdot\frac{du_{\lambda,jn}(x)}{dx}\right) \tag{6.7}$$

From this is evident that kinetic energy density should be constant over any region where $V_\lambda(x)=0$. This is also the case for the solution $T_{00,R}\left(x;|0_\lambda\rangle\right)$ as determined by mode renormalization. It is constant over Region I and Region II. Next determine the change in $T_{00,R}\left(x;|0_\lambda\rangle\right)$ in going from Region II to Region I. In this case we obtain,

$$\Delta T_{00,R}\left(II \to I; |0_\lambda\rangle\right) = \lambda\int_{x_1}^{x_2}\sum_{jn}\left(\delta(x+a/2)u_{\lambda,jn}(x)\cdot\frac{du_{\lambda,jn}(x)}{dx}\right)dx \tag{6.8}$$

where $x_1 < -a/2$ and $a/2 > x_2 > -a/2$. One difficulty in evaluating this is that the first derivative of $u_{\lambda,jn}(x)$ is not continuous at the boundary $x = -a/2$. To deal with this we write the following expression for this first derivative which is valid in the vicinity of $x = -a/2$,

$$\frac{du_{\lambda,jn}(x)}{dx} = \frac{du_{\lambda,jn}(x;II)}{dx}\theta\left(-\left(x+\frac{a}{2}\right)\right) + \frac{du_{\lambda,jn}(x;I)}{dx}\theta\left(x+\frac{a}{2}\right) \quad (6.9)$$

Also use $\delta(x) = d\theta(x)/dx$ and $\delta(x) = -d\theta(-x)/dx$. Also note that $\theta(x)^2 = \theta(x)$. This leads to the following relationships,

$$\theta(x)\delta(x) = \frac{1}{2}\frac{d\theta(x)^2}{dx} = \frac{1}{2}\frac{d\theta(x)}{dx} = \frac{1}{2}\delta(x) \quad (6.10)$$

and

$$\theta(-x)\delta(x) = -\frac{1}{2}\frac{d\theta(-x)^2}{dx} = \frac{1}{2}\delta(x) \quad (6.11)$$

From all this we obtain,

$$\Delta T_{00,R}\left(II \to I; |0_{\lambda}\rangle\right) = \frac{\lambda}{2}\int_{x_1}^{x_2}\sum_{jn}\left(\delta(x+a/2)u_{\lambda,jn}(x)\cdot\left(\frac{du_{\lambda,jn}(x;II)}{dx} + \frac{du_{\lambda,jn}(x;I)}{dx}\right)\right)dx \quad (6.12)$$

This becomes,

$$\Delta T_{00,R}\left(II \to I; |0_{\lambda}\rangle\right) = \sum_{jn}\Delta T_{jn} \quad (6.13)$$

where,

$$\Delta T_{jn} = \frac{\lambda}{4}\left(u_{\lambda,jn}(-a/2)\cdot\left(\frac{du_{\lambda,jn}(-a/2;II)}{dx} + \frac{du_{\lambda,jn}(-a/2;I)}{dx}\right)\right) \quad (6.14)$$

Use (5.5) along with (5.6) to obtain,

$$\Delta T_{1n} = -\frac{\lambda N_{1n}^2 A_{1n}^2}{2L\omega_{\lambda,1n}}\left(\frac{2\omega_{\lambda,1n}\sin(\omega_{\lambda,1n}a/2)\cos(\omega_{\lambda,1n}a/2)}{+\lambda\sin^2(\omega_{\lambda,1n}a/2)}\right) \quad (6.15)$$

This in turn equals,

$$\Delta T_{1n} = \frac{\omega_{\lambda,1n}N_{1n}^2}{2L}\left(A_{1n}^2 - 1\right) \quad (6.16)$$

Similarly, for $\Delta T_{2n}$ we obtain,

$$\Delta T_{2n} = \frac{\omega_{\lambda,2n}N_{2n}^2}{2L}\left(A_{2n}^2 - 1\right) \quad (6.17)$$

Use these results in (6.13) to obtain,

$$\Delta T_{00,R}\left(II \to I; |0_\lambda\rangle\right) = \sum_{jn} \frac{N_{jn}^2 \omega_{\lambda,jn}}{2L}\left(A_{jn}^2 - 1\right) \qquad (6.18)$$

This expression is equivalent to,

$$\Delta T_{00,R}\left(II \to I; |0_\lambda\rangle\right) = \frac{1}{2L}\sum_{jn}\left\{\begin{array}{l}\left(N_{jn}^2 \omega_{\lambda,jn} A_{jn}^2 - \omega_{0,jn}\right) \\ -\left(N_{jn}^2 \omega_{\lambda,jn} - \omega_{0,jn}\right)\end{array}\right\} \qquad (6.19)$$

Compare this result to (5.16) and (5.17) to obtain,

$$\Delta T_{00,R}\left(II \to I; |0_\lambda\rangle\right) = T_{00,R}\left(I; |0_\lambda\rangle\right) - T_{00,R}\left(II; |0_\lambda\rangle\right) \qquad (6.20)$$

Therefore the expression we obtain in Section 5 for the kinetic energy density using mode renormalization is consistent with the results obtained in this section using a renormalization free procedure.

**7. Conclusion.**

We have defined a state vector $|0_\lambda\rangle$. Using the results from [11] we have calculated the kinetic energy density of this state vector using mode renormalization. We have shown that this kinetic energy density violates the spatial quantum inequality. This is consistent with the results of [8] and [9] which also demonstrate a violation of the spatial quantum inequality.

**References.**

1. H. Epstein, V. Glaser, and A. Jaffe, *Nonpositivity of the energy density in quantized field theories*, Nuovo Cim. **36** (1965) 1016-1022.

2. L.H. Ford and T.A. Roman, *Restrictions on Negative Energy Density in Flat Spacetime*, Phys. Rev. **D55** (1997) 2082-2089. (Also axXiv:gr-qc/9607003).

3. E.E. Flanagan, *Quantum inequalities in two dimensional Minkowski spacetime*, Phys. Rev. **D56** (1997) 4922-4926. (Also arXiv:gr-qc/9706006).

4. C.J. Fewster and S.P. Eveson, *Bounds on negative energy densities in flat spacetime,* Phys. Rev. **D58**, 084010, 1998. (Also arXiv:gr-qc/9805024).

5. L.H. Ford, *Negative Energy Densities in Quantum Field Theory,* arXiv:0911.3597.